\documentclass[a4paper,twocolumn,showpacs,nofootinbib,floatfix,prc]{revtex4}

\usepackage{graphicx}
\usepackage{amsfonts}
\renewcommand{\vec}[1]{\mathbf{#1}}
\def\be{\begin{equation}}
\def\ee{\end{equation}}
\def\bea{\begin{eqnarray}}
\def\eea{\end{eqnarray}}
\newcommand{\beq}{\begin{equation}}
\newcommand{\eeq}[1]{\label{#1} \end{equation}} 
 
\begin{document}

\title{\Large \bf Critical review of \boldmath{$[K^-ppn]$} bound states}

\author{\large V.K. Magas$^1$,  E. Oset$^2$, A. Ramos$^1$} 

\affiliation{
 $^1$~Departament d'Estructura i Constituents de la Mat\`eria\\
Universitat de Barcelona,  Diagonal 647, 08028 Barcelona, Spain \\  
 $^2$~ Departamento de F\'{\i}sica Te\'orica and \\
IFIC Centro Mixto Universidad de Valencia-CSIC\\
Institutos de Investigaci\'on de Paterna,\\
Apdo. correos 22085, 46071, Valencia, Spain
}

\begin{abstract}
We make a thorough study of the process of three body kaon
absorption in nuclei, in connection with a recent FINUDA experiment which
claims the existence of a deeply bound kaonic state from the observation of 
a peak in the 
$\Lambda d$ invariant mass distribution following $K^-$ absorption on $^6$Li. 
We show that the peak is naturally
explained in terms of $K^-$ absorption from three nucleons leaving the rest as
spectators. We can also reproduce all the other observables measured in the same
experiment and used to support the hypothesis of the deeply bound kaon state.
Our study also reveals interesting aspects of kaon absorption in nuclei, a 
process that must be understood in order to make 
progress in the search for $K^-$ deeply bound states in nuclei. 
\end{abstract}


\maketitle

\section{Introduction}

The possibility of having deeply bound $K^-$ states in nuclei is drawing much
attention both theoretically and experimentally. The starting point to face
this problem is obviously the understanding of the elementary $\bar{K}N$
interaction, and lots of efforts have been devoted to this topic, mostly  using
unitary extensions of chiral perturbation theory 
\cite{weise,angels,Oller:2000fj,Oset:2001cn,jido,Garcia-Recio:2002td,Garcia-Recio:2003ks,Hyodo:2002pk,GarciaRecio:2005hy,Hyodo:2007jq}. 
The recent determination of the $K^- p$ scattering  length from  the study of
$K^- p$ atoms in DEAR at DA$\Phi$NE \cite{Beer:2005qi} has stimulated  a
revival of the interest on this issue and several studies have already
incorporated  chiral Lagrangians  of higher order
\cite{Borasoy:2005ie,Oller:2005ig,Oller:2006jw,borasoy} in addition to the
lowest order one used in  \cite{angels,Oller:2000fj,Oset:2001cn}.

 Much work has also been done  along these lines in order to study the
interaction of kaons with nuclei, deducing $K^-$ nucleus optical potentials 
with a moderate attraction of about 50 MeV at normal nuclear matter density 
\cite{lutz,angelsself,schaffner,galself,Tolos:2006ny}. The selfconsistency of
the calculation is an important requirement for the construction of the
potential, due to the presence of the $\Lambda(1405)$ resonance below
threshold, and is responsible for a fast transition from a repulsive potential
in  the $t \rho$ approximation at very low densities to an attraction at the
densities felt by measured  kaonic atom states. This ``shallow"  theoretical
potential was shown to reproduce satisfactorily the data on shifts and widths
of kaonic atoms \cite{okumura}. However, reduced chi-squared values were obtained from
phenomenological fits to kaonic
atoms which favored strongly attractive potentials of the order of $-200$ MeV
 at the center of the nucleus
\cite{friedman-gal}. A combination of theory and phenomenology was attempted in
Ref.~\cite{baca}, where an excellent
fit to the full set of kaonic atom data was found with a potential 
that deviated at most by 20\% from the theoretical one of \cite{angelsself}.  This
potential also generated deeply bound  $K^-$ nuclear states having a width of
the order of 100 MeV, much bigger than the corresponding binding energy. The
bound states would then overlap among themselves and with the continuum and,
consequently, would not show up as narrow peaks in an experiment. 

Other phenomenological potentials of sizable attraction
(with potential depths around 100--200 MeV at $\rho_0$) 
that could in principle 
accommodate deeply bound states, have been discussed in 
\cite{gal1,gal2,gal3,gal4,gal,muto,amigo1,amigo2}. 
In these latter works a relativistic mean field
approach is followed, introducing $\sigma$ and $\omega$ fields which couple to kaons and
nucleons to obtain the $\bar{K}$ nucleus optical potential. Less attractive potentials
within this framework are also found in
\cite{Zhong:2004wa,Zhong:2006hd,Dang:2007ai}. A new look at these
relativistic mean field potentials from the perspective of the microscopic 
chiral unitary approach is presented in \cite{Torres:2007rz}.

    The opposite extreme has been represented by some highly 
    attractive phenomenological potentials with about
600 MeV strength in the center of the nucleus \cite{akaishi,akainew}.  These
potentials, leading to compressed
nuclear matter of ten times nuclear matter density, met criticisms from 
\cite{toki}  and more 
recently from  \cite{Hyodo:2007jq}. The work of \cite{toki} met criticism from 
\cite{akanuc} concerning the "narrow peak" predicted in  \cite{toki}, but actually
the width of the peak was not calculated in \cite{toki}. It was calculated in 
\cite{hyper} showing that it was not narrow and indeed agrees with the revised
experiment of the KEK work as we shall discuss below. The criticisms of
\cite{akanuc} were rebutted in \cite{hyper} and more recently in
\cite{npangels}.

Predictions of deeply bound $K^-$ states for few nucleon systems have been 
first done
in \cite{akaishi,akainew}. More recently, 
Faddeev-type calculations were made for the $\bar{K}NN$ system
using phenomenological input in \cite{Shevchenko:2006xy,Shevchenko:2007zz}
and a leading-order
chiral interaction in \cite{Ikeda:2007nz}. Both studies found a $K^-pp$ quasibound
state above the $\pi \Sigma N$ threshold with a relatively
large width. A variational approach with phenomenological
local potentials has also been applied in \cite{Yamazaki:2007cs} to study the 
$\bar{K} NN$ system, leading to a bound state at
about 50 MeV below the $\bar{K} NN$ threshold. A more recent 
variational calculation  \cite{Weise:2007rd,Dote:2007rk}
emphasizes the important role of the
repulsive $NN$ interaction at short distances and obtains preliminary results
having smaller bindings and larger widths than those found by 
the other earlier approaches.

On the experimental side the situation is still at a very early stage. Initial hopes
that a peak seen in the  $(K^-_{stop},p)$ reaction  on $^4$He
\cite{Suzuki:2004ep} could be a signal of a $K^-$ bound in the trinucleon with
a binding of 195 MeV gra\-dua\-lly lost a support. First, an alternative
explanation of the peak was presented in \cite{toki}, 
showing that a peak with the strength claimed in the experiment
was coming from $K^-$ absorption on a pair of nucleons going to $p \Sigma$, 
leaving the other two nucleons
as spectators. This hypothesis led to the prediction that such a peak should
be seen in other light or medium nuclei where it should be narrower and weaker
as the nuclear size increases. This was
confirmed with the finding of such a peak in the $(K^-_{stop},p)$ reaction on
$^6$Li, which already fades away in $^{12}$C nuclei at FINUDA 
\cite{agnellonuc}. In \cite{toki} the $K^-$ absorption was described as taking place from 
 (np) pairs of the Fermi sea. In \cite{agnellonuc} the same explanation was
given for the peaks suggesting that
the (np) pairs would be correlated in "quasi"-deuteron clusters. 
The final development in this discussion
 has come from a new experiment of the KEK reaction of \cite{Suzuki:2004ep} 
 reported in 
 \cite{Sato:2007sb} where,
 performing a more precise measurement, which subsaned deficiencies in the efficiency
 corrections, 
 the relatively narrow
 peak seen in \cite{Suzuki:2004ep} disappears
 and only a  broad bump remains around the region where the peak was
 initially claimed. The position and width of this broad bump are in
 agreement with the estimations done in \cite{hyper,npangels} based on the 
 kaon absorption mechanism of \cite{toki}. 
 
    The second source of initial hope came from the experiment of the FINUDA
    collaboration 
    \cite{Agnello:2005qj}, where a peak seen in the invariant mass distribution
    of $\Lambda p$ following $K^-$ absorption in a mixture of light nuclei was
    interpreted as evidence for a $K^- pp$ bound state, with 115 MeV binding and
    67 MeV width. However, it was shown in \cite{Magas:2006fn,Kpp} that the peak
    seen could be interpreted in terms of $K^-$ absorption on a pair of nucleons
    leading to a $\Lambda p$ pair, followed by the rescattering of the $p$ or the
    $\Lambda$ on the remnant nucleus. 
    
    More recently, a new experiment of the FINUDA collaboration \cite{:2007ph}
    found a peak on the invariant mass of $\Lambda d$ following the absorption
    of a $K^-$ on $^6$Li, which was interpreted as a signature for a bound
    $\bar{K}NNN$ state with 58 MeV binding and 37
    MeV width. These results are puzzling, since 
    the bound state of the ${\bar K}$ in the three nucleon system has 
    significantly smaller values for the binding and width than those claimed
    for the bound state of the ${\bar K}$ in the two nucleon system 
    \cite{Agnello:2005qj}.
    These unexpected results require serious
    thoughts but no discussion was done in \cite{:2007ph}.  
    
    About the same time as the FINUDA experiment \cite{:2007ph}
    a similar  experiment was performed at KEK 
    \cite{Suzuki:2007kn}, looking also at the $\Lambda d $ invariant mass 
    following $K ^-$ absorption but on a $^4$He target. The authors of this 
    latter work do not
    share the conclusions of  \cite{:2007ph} concerning the association of the
    peak to a $\bar{K}$ bound state, and claim instead that the peak could be
     a signature of three body absorption.  
    
       In the present work we perform
       detailed calculations of $K^-$ absorption from three nucleons in $^6$Li
       and show that
       all features observed in the experiment of \cite{:2007ph} can be well
       interpreted in the picture of three body kaon absorption, as suggested in
    \cite{Suzuki:2007kn}, with the rest of the nucleons acting as
    spectators.

\section{Mechanism for $K^-$ three body absorption}

In the $K^{-}_{stop} A \to \Lambda d A'$ reaction \cite{:2007ph},  at least
three nucleons must participate in the absorption process.  Two body $K^-$
absorption processes of the type $K^- NN \to \Sigma N ( \Lambda N) $  have been
studied experimentally in \cite{Katz:1970ng} and their strength is seen to be
smaller than that of the one body absorption  $K^- N \to \pi \Sigma (\pi
\Lambda )$ mechanisms. This result follows the  argument that it is easier to
find one nucleon than two nucleons together in the nucleus. This is also the
case in pion absorption in nuclei, where extensive studies, both theoretical
\cite{Oset:1986yi} and experimental \cite{Weyer:1990ye},  obtain the direct two
and three body absorption rates with the former one dominating over the later,
particularly for pions of low  energy. We follow here the same logics and
assume the process to be dominated by direct three body $K^-$ absorption, the
four body playing a minor role. 

   The former assumption means in practice that the other three nucleons not
directly involved in the absorption process will be spectators in the reaction.
  These three spectator nucleons have to leave the nucleus, but
  they were bound in $^6$Li. The nuclear dynamics takes care of this since 
  there is a distribution of momenta and energies in the nucleus, and the
  ejection of either three nucleons, a $n d$  pair or tritium, implies that the
  absorption is done in the most bound nucleons. 
  
  The other element of relevance
  is the atomic orbit from which the kaon is absorbed. This information is 
  provided
  by the last measured transition in the X-ray spectroscopy of $K^-$-atoms, 
  which occurs precisely
  because absorption overcomes the $\gamma$ ray emission. In the case of $^6$Li
  this happens for the $2p$
  atomic state \cite{okumura,friedman-gal}.  
  
  Following the line of studies done for pion absorption 
  and other inclusive reactions \cite{Salcedo:1987md}, 
 we describe the nucleus in terms of a local Fermi sea with Fermi momentum
 $k_F(r)$. The nucleons move in a mean field
given by the Thomas Fermi potential
 \beq
V(r)=-\frac{k_F^2(r)}{2m_N}\,, \quad k_F(r)=\left(\frac{3\pi^2}{2}\rho(r) \right)^{1/3}\,,
\eeq{e1}
 where $m_N$ is the nucleon mass and $\rho(r)$ is the local nucleon density 
 inside the nucleus. 
 
 This potential assumes a continuity from the energies
 of the bound states (holes) to those in  the continuum (particles), which is
 not the case in real nuclei. For this reason, we implement an energy gap, 
 $\Delta$, which is adjusted to respect the threshold of
 the reaction. The introduction of a gap in the Fermi sea is a common practice 
 in order to be precise with the actual binding
 energies of the  nuclei involved in a particular reaction so that the corresponding threshold is respected 
 \cite{Kosmas:1996fh,Albertus:2001pb,Nieves:2004wx}.
 Hence, we demand that the highest possible invariant mass of 
  $K^- NNN$ system, which happens when the three nucleons are at the
 Fermi surface with total three-momentum zero, corresponds to the minimum
 possible energy for a spectator three-nucleon system with total zero momentum,
 namely a tritium at rest. 
   This situation corresponds to
  \beq
  m_{K^-} + M_{ ^6{\rm Li}}= 
  m_{K^-}+3m_N-3\Delta+M_t \ ,
  \eeq{e2}
 and we determine $\Delta=7.8$ MeV. In the above expression $m_{K^-}$,  $M_t$,
 $M_{ ^6{\rm Li}}$ are the masses of the 
 corresponding particles and nuclei. 
   
   The probability of $K^-$ absorption by three nucleons will be determined from
   the third power of the nuclear density as
  \beq
 \Gamma \propto \int d^3 \vec{r}  |\Psi_{K^-}(r)|^2 \rho^3(r)\,,
  \eeq{e3}
  where $\Psi_{K^-}(r)$ is the $K^-$ atomic wave function.
   In order to take into account the Fermi motion we write the density as 
$\rho(r)=4 \int \frac{d^3 \vec{p}}{(2\pi)^3} \Theta(k_F(r)-|\vec{p}|)$ 
   and then we obtain
$$ 
\Gamma \propto \int d^3 \vec{r} d^3 \vec{p}_1  d^3 \vec{p}_2 d^3 \vec{p}_3 |\Psi_{K^-}(r)|^2\times
$$
\beq
 \times \Theta(k_F(r)-|\vec{p}_1|)\Theta(k_F(r)-|\vec{p}_2|)\Theta(k_F(r)-|\vec{p}_3|)\,.  
\eeq{P}

   From this expression we can evaluate all observables of the reaction. Let us first
   concentrate on the $\Lambda d$ invariant mass which, for each $K^- NNN \to
   \Lambda d$ decay event, is precisely the invariant mass of the 
   corresponding $K^-NNN$ system, the other three nucleons acting as spectators.
   Thus the energy of the  $\Lambda d$ pair is obtained from
   \begin{eqnarray}
 &&E_{\Lambda d} = E_{K^-NNN} \equiv E_{K^-}+ E_{N_1} + E_{N_2} + E_{N_3}\\
   &&\!\!\!\! =
   \!m_{K^-}\!+\!3m_N\! +\! \frac{\vec{p}_1^2}{2m_N}\!+\!
   \frac{\vec{p}_2^2}{2m_N}\!+\!
   \frac{\vec{p}_3^2}{2m_N}\!-\!3\frac{k_F^2(r)}{2m_N}\!-\!3\Delta \ , \nonumber 
   \label{e6a}
 \end{eqnarray}
 and the momentum from
   \beq
   \vec{P}_{\Lambda d}=\vec{P}_{K^-NNN} =
   \vec{p}_1+\vec{p}_2+\vec{p}_3 \ ,
   \eeq{e6b}
and, correspondingly,
   \beq
   M_{\Lambda d} = E_{\Lambda d} - \frac{\vec{P}_{\Lambda d}^2}{2E_{\Lambda d}}\ .
   \eeq{e6}
   One may also easily obtain the invariant mass of the residual system,
   $M^*$, from
  \begin{equation}
     M^* = E^* - \frac{\vec{P}^{*\,2}}{2E^*}\ ,
     \label{mst}
  \end{equation}
  with
   \beq
   E^*=m_{K^-}+M_{ ^6{\rm Li}}-E_{K^-NNN}\ , \quad \vec{P}^*=-\vec{P}_{\Lambda d}\ .
    \eeq{aa}

  Each event in the multiple integral of Eq.~(\ref{P}), done with the Monte Carlo
  method, selects particular values for $\vec{r}$, $\vec{p}_1$, $\vec{p}_2$ and 
  $\vec{p}_3$ which, in turn, determine the value of the corresponding $\Lambda d$ 
  invariant mass from Eqs.~(\ref{e6a})--(\ref{e6}). 
   Since the minimum obvious invariant mass of the residual
     three-nucleon system is $M^*=M_t$, corresponding to the emission of
     tritium, the cut $\Theta(M^* - M_t)$ is also imposed for each event. 
  A compilation of events provides us with
   the $\Lambda d$ invariant mass distribution.    
   We also directly obtain the distribution of
   total $\Lambda d$ momentum, Eq.~(\ref{e6b}), to be directly compared with 
   the $\Lambda d$ momentum measured in \cite{:2007ph}. 
   
   Please note that the model presented here is a straightforward generalization (from two nucleon to three nucleon $K^-$ absorption)
    of the one used in Refs. \cite{Magas:2006fn,Kpp}, however here we concentrate on the primary reaction peak, while in Refs. \cite{Magas:2006fn,Kpp}     the authors were more interested in the peak generated by the final state interactions, i.e. by the collisions of the primary produced $\Lambda$     and $p$ on their way out of the nucleus. Since the two nucleon $K^-$ absorption, discussed in \cite{Magas:2006fn,Kpp}, was measured for heavier    nuclei \cite{Agnello:2005qj} the final state interaction peak was stronger than that of the primary reaction, contrary to the reaction studied in       this work. 
    
   Other observables measured in \cite{:2007ph} require an additional work.
   One is the angular
   correlation of $\Lambda d$ pairs, and the other is the missing mass 
   assuming a residual $n d$ system, apart from the measured $\Lambda d$ pair, 
   namely
   \beq
   T_{miss} = m_{K^-}+M_{ ^{6}{\rm Li}} - m_\Lambda - m_n - 2 M_d - (T_\Lambda + T_d) 
   \ ,
   \eeq{e7}
   where $m_\Lambda$, $M_d$ and $T_{\Lambda}$, $T_d$ are the masses and the 
   kinetic energies of the  $\Lambda$ and the $d$, correspondingly. These two
   observables require the evaluation of the individual $\Lambda$ and $d$
   momenta in the laboratory frame. Their value in 
   the center of mass (CM) frame of the $\Lambda d$ pair is given
   in terms of the known invariant mass but their direction in this frame is
   arbitrary. We take this into account by obtaining $\Lambda$ and $d$ momenta
   in the CM frame
     \begin{eqnarray}
  \vec{p}_\Lambda^{CM}&=&p_{\Lambda}^{CM}\left(\sin\Theta \cos\phi, 
  \sin\Theta \sin\phi, \cos\Theta \right)\ , \nonumber \\
   \vec{p}_d^{CM}&=&-\vec{p}_\Lambda^{CM} \ ,
  \end{eqnarray}
  with
    \beq
  p_\Lambda^{CM}=\frac{\lambda^{1/2}(M_{\Lambda
  d}^2,m_{\Lambda}^2,M_d^2)}{2M_{\Lambda d}}\ , 
  \eeq{e8}
  where the events are now generated according to the distribution provided by
  the integral 
\begin{eqnarray}
&& \int d \cos\Theta \int d\phi \int d^3 \vec{r} d^3 \vec{p}_1  d^3 \vec{p}_2 d^3 \vec{p}_3 
 |\Psi_{K^-}(r)|^2 \nonumber \\ 
&& \times \Theta(k_F(r)-|\vec{p}_1|)
 \Theta(k_F(r)-|\vec{p}_2|)\Theta(k_F(r)-|\vec{p}_3|)\nonumber \\
 && \times \Theta(M^* - M_t) \ .
  \end{eqnarray}
In order to have the final $\Lambda$ and $d$ momenta in the laboratory frame,
where the $\Lambda d$ pair has momentum $\vec{P}_{\Lambda d}$, we apply the 
transformations
  \begin{eqnarray}
  \vec{p}_\Lambda &=&\vec{p}_\Lambda^{CM}+ m_\Lambda \vec{v} \nonumber \\
 \vec{p}_d&=& -\vec{p}_\Lambda^{CM}+ M_d \vec{v}\ ,
  \end{eqnarray}
  where $\vec{v}=\vec{P}_{\Lambda d}/(m_\Lambda+M_d)$.
  These last equations allow us to find the cosinus of the angle between 
  the directions of $\Lambda $ and $d$. Therefore, generating the 
  distribution of events according to their relative angle is
  straightforward. We will see, as it is also the case of the experiment,
  that  $P_{\Lambda d}\sim 200$ MeV/c, while $p_{\Lambda}^{CM} \sim 650$ 
  MeV/c, which already guarantees that the $\Lambda d$ events will be largely 
  correlated back-to-back.
   
    We note that our calculations incorporate the same momentum cuts as
    in the experiment, namely  140  MeV/c $< p_{\Lambda}<$ 700 MeV/c
 and  300  MeV/c $< p_d <$ 800  MeV/c.

\begin{figure}[ht]
\vspace{-0.0cm}
\begin{center}
\includegraphics[width=8.4cm]{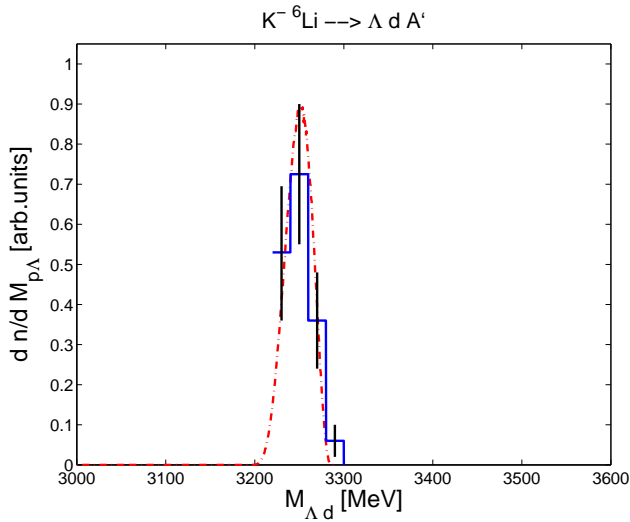}
\end{center}
\vspace{-0.0cm}
 
\caption{(Color online) The $\Lambda d$ invariant mass distribution for the $K^{-}_{stop} A
\to \Lambda d A'$ reaction. Histogram and error bars are from the experimental
paper \cite{:2007ph}, while the dot-dashed curve is the result of our calculation.}

\label{fig1}
\vspace{-0.0cm}
\end{figure}

 \section{Results}
 
 In Fig. \ref{fig1} we show the results for the invariant mass of the $\Lambda d$ system.
 Our distribution, displayed with a dot-dashed line, peaks around 
 $M_{\Lambda d}= 3252$ MeV as in the experiment. The shape of the distribution
 also compares remarkably well with the experimental histogram in the region of
 the peak, which is the energy range that we are exploring in the present work. 
 We obtain a width of about $36$ MeV, as reported in the
 experiment. Note that apart from the peak that we are discussing, the
 experiment also finds events at lower $\Lambda d$ invariant masses which did
 not play a role in their discussion \cite{:2007ph}. These
 events would be generated in cases where there is final state interaction of
 the $\Lambda$ or the $d$  with the
 rest of the nucleons, as was discussed in \cite{Magas:2006fn,Kpp}, or through 
 other absorption mechanisms, but this is not the 
 object of discussion here, as well as in \cite{:2007ph}. 
 
\begin{figure}[ht]
\vspace{-0.0cm}
\begin{center}
\includegraphics[width=8.4cm]{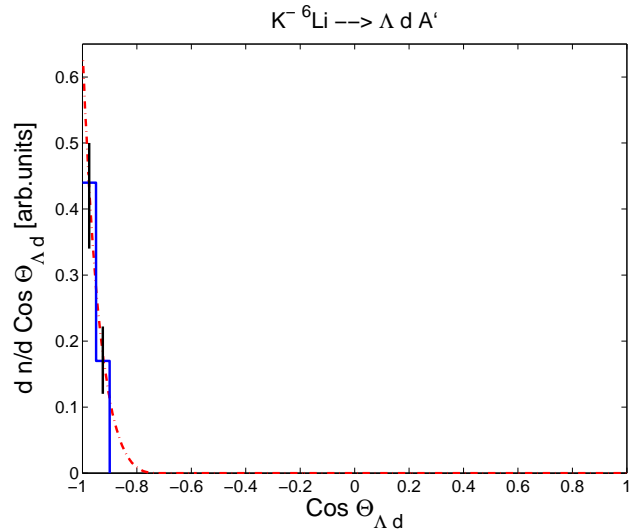}
\end{center}
\vspace{-0.0cm}
\caption{(Color online) The $\Lambda d$ angular distribution for the $K^{-}_{stop} A \to
\Lambda d A'$ reaction. Histogram and error bars are from the experiment
\cite{:2007ph}, while the dot-dashed curve is the result of our calculation. 
As in the experimental analysis, we take into account the following cuts: 
 3220 MeV $ < M_{\Lambda d} <$ 3280  MeV.}
\label{fig2}
\vspace{-0.0cm}
\end{figure}

  The angular correlations between the emitted $\Lambda$ and $d$ can be seen in
  the distribution displayed in Fig. \ref{fig2}, where, as
  in the experimental analysis, we consider only those events
  which fall  in the region  3220  MeV $< M_{\Lambda d} <$ 3280  MeV.
  As we can see in the figure, the
  distribution is strongly peaked backward and the agreement with experiment is
  very good.

\begin{figure}[ht]
\vspace{-0.0cm}
\begin{center}
\includegraphics[width=8.4cm]{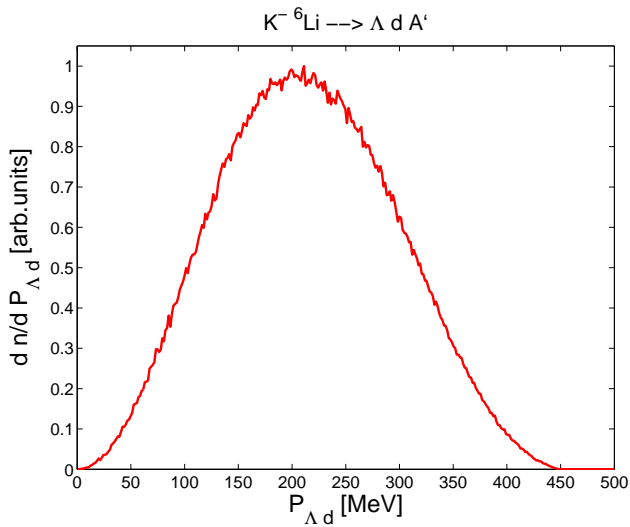}
\end{center}
\vspace{-0.0cm}
\caption{(Color online) The $\Lambda d$ momentum distribution for the $K^{-}_{stop} A \to
\Lambda d A'$ reaction. The calculation implements the cut
3220 MeV $< M_{\Lambda d} <$ 3280 MeV.}
\label{fig3}
\vspace{-0.0cm}
\end{figure}

    The distribution of the total $\Lambda d$ momentum in the mass
    range of the bump is shown in Fig. \ref{fig3}. The experimental paper does
    not show a distribution but quotes that it peaks around 190 MeV/c, which 
    is precisely the region where the peak of our calculated spectrum lies.

\begin{figure}[ht]
\vspace{-0.0cm}
\begin{center}
\includegraphics[width=8.4cm]{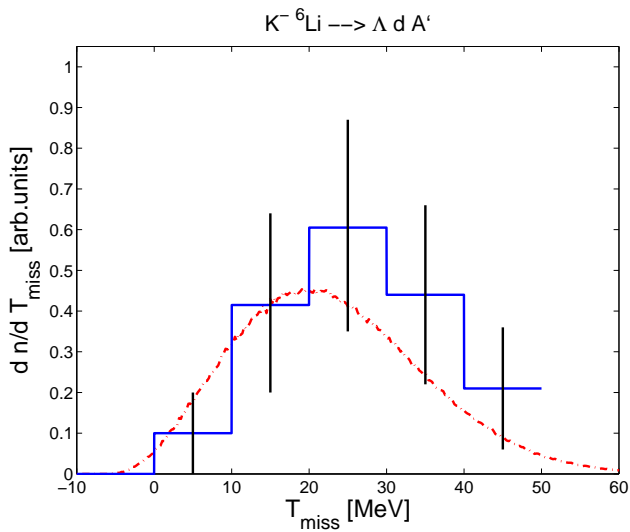}
\end{center}
\vspace{-0.0cm}
\caption{(Color online) The missing mass distribution for the $K^{-}_{stop} A \to \Lambda d
A'$ reaction. Histogram and error bars are from the experimental paper
\cite{:2007ph}, the full curve is the result of our calculation.}
\label{fig4}
\vspace{-0.0cm}
\end{figure}

    Our results for the missing mass distribution, defined by Eq.~(\ref{e7}),
    are compared with the experimental data in Fig. \ref{fig4}.  As we can see,
    the agreement with experiment is reasonably good within the large experimental errors.
    We should remark here that the peak in Fig. \ref{fig4} was
    associated in \cite{:2007ph} to the mechanism of $K^-$ absorption in a
     $^4$He cluster, namely $K^- + \alpha (d) \to \Lambda d n (d)$,
    motivated by the assumption that the $^6$Li nucleus  is largely made of a  $\alpha$
    particle and a deuteron. As we can see, our approach, which relies upon
    three body absorption, reproduces the data which, thus, cannot be
    taken as evidence for the mechanism claimed in \cite{:2007ph}.

\begin{figure}[ht]
\vspace{-0.0cm}
\begin{center}
\includegraphics[width=8.4cm]{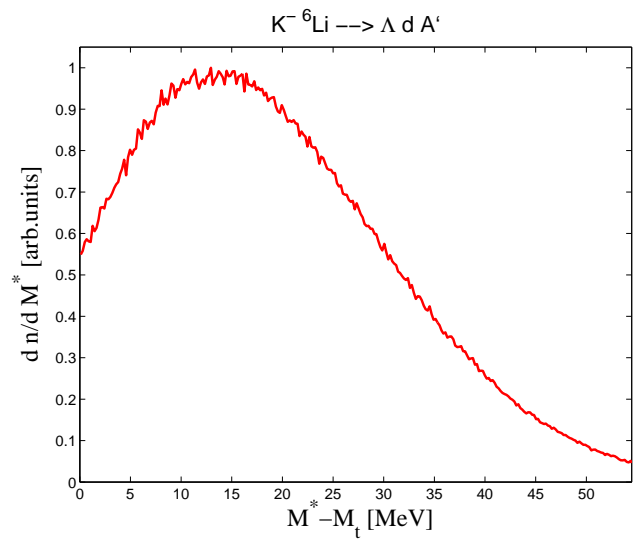}
\end{center}
\vspace{-0.0cm}
\caption{(Color online) The invariant mass distribution of the residual system, 
Eq.~(\protect\ref{mst}), for the $K^{-}_{stop} A \to \Lambda d A'$ reaction. }
\label{fig5}
\vspace{-0.0cm}
\end{figure}
    
    An alternative way to present this information in a way more closely 
   related to the mechanism we have, is by looking at the invariant mass
   distribution of the residual three particle state (the spectator nucleons in
   our case). This is shown in Fig.~\ref{fig5}, where the invariant mass is
   measured with respect to $M_t$, a natural threshold which is imposed in our
   formalism. We observe a peak in the mass distribution at energies around
   $20$ MeV, clearly higher than the tritium binding energy of $8.48$ MeV. 
   This means that there is not only room for $t$ production, but also for $d
   n$ production, as assumed to be the case in \cite{:2007ph}, and for
   uncorrelated three nucleon emission. 

\begin{figure}[ht]
\vspace{-0.0cm}
\begin{center}
\includegraphics[width=8.4cm]{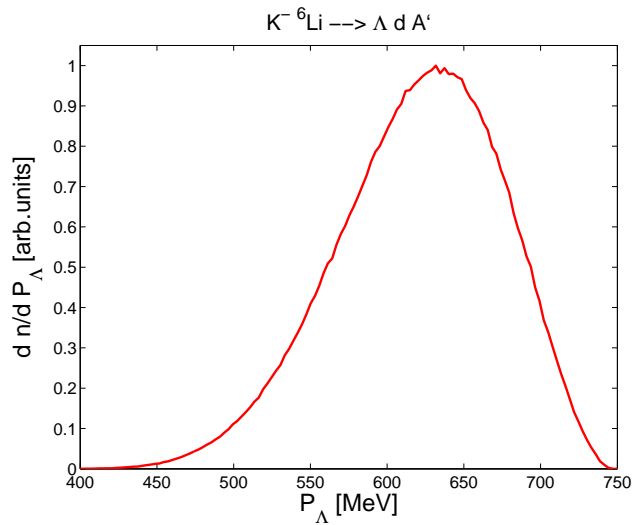}
\end{center}
\vspace{-0.0cm}
\caption{(Color online) The $p_\Lambda$ distribution for the $K^{-}_{stop} A \to \Lambda d A'$ reaction.}
\label{fig6}
\vspace{-0.0cm}
\end{figure}

      In Fig.~\ref{fig6} we present the momentum distribution of the $\Lambda$,
which we can also compare with the experimental
observations. For the results shown in the figure we removed the 
momentum cuts, 140  MeV/c $< p_{\Lambda}<$700 MeV/c and   300  MeV/c $< p_d <$
800 MeV/c. We observe that the $\Lambda$ momentum peaks around 635 MeV/c and
most of the events are contained in the region between 450  MeV/c and 700 MeV/c,
as also found in the experiment. All our events are contained within the
experimental window for $p_d$ momentum.

\section{Empirical qualitative discussion of the strength of the reaction}

   In addition to the observables discussed in the previous section the yield of
the  observed peak is also given in \cite{:2007ph} as 
$Y_{\Lambda d}~=~(4.4\pm1.4) \times 10^{-3}/K^-_{stop}$.
In the former discussion we did not use a specific dynamical model for the 
absorption of the $K^-$ by three nucleons. This is, we did not use specific
Lagrangians and a set of Feynman diagrams which would have given us the strength
of the absorption process. 
In experimental studies the yield is simply the number of events for a particular 
channel per stopped  $K^-$. 
In contrast a theoretical determination of the yield of a process, 
or in other words the fraction of the total rate that goes into a particular channel, 
requires the calculation of all possible reaction processes. 
Clearly this is a very hard and time demanding task. Indeed, the
experimental work of $K^-$ absorption on $^4He$ \cite{Katz:1970ng} quotes
in Table III a list of 21 reactions following $K^-$ absorption by one nucleon 
(mesonic) and multinucleon (nonmesonic) mechanisms, and this represents only a fraction
of the total. In many cases, one has nuclei in the final state which complicates further 
an eventual theoretical calculation. The present theoretical situation
is such that the microscopic mechanisms for two nucleon kaon absorption are only
available from \cite{angelsself,schaffner}. There is no work done on three nucleon 
$K^-$ absorption, and the scarce theoretical work on three nucleon absorption of 
pions \cite{Oset:1986yi,harry} is a reflection of the intrinsic theoretical
difficulties that any microscopic evaluation involves. This, together with the
enormous amount of physical channels that one would have to evaluate to produce
the relative yield of just one of them, describes clearly the horizon of such a goal.

   In view of this horizon the work presented here becomes even more valuable, 
because it has demonstrated that the observables presented in \cite{:2007ph} to
support the idea of a kaon bound state could be reproduced with just the
kinematics of the three body absorption mechanism, and the detailed dynamical
mechanisms that would have allowed us an evaluation of the absolute strength of
the reaction were never needed. 
We also note that the yield of the peak was not offered  as a proof for the 
advocated $K^ -$ bound state in \cite{:2007ph}, since 
the strength itself provides no information on the mass and width which are the
characteristics of a physical state. 

However, we shall make some instructive discussion about this
yield from the empirical point of view with the only
purpose to gain some knowledge on $K^-$ absorption.
This will illustrate that experiments like the one we are discussing provide 
indeed valuable information on $K^-$ absorption worth giving some thoughts to. 

\begin{figure}[ht]
\vspace{-0.0cm}
\begin{center}
\includegraphics[width=8.4cm]{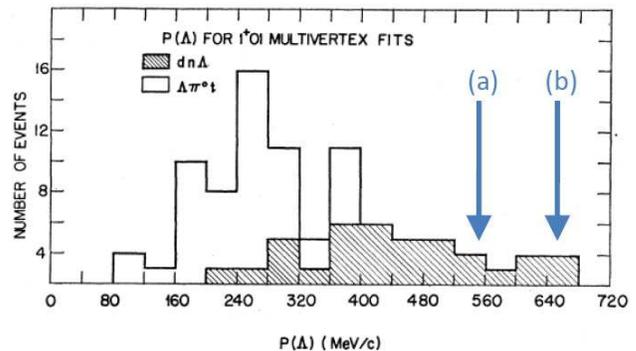}
\end{center}
\vspace{-0.5cm}
\caption{(Color online) The $\Lambda$ momentum distribution for $K^-$ absorption  
reactions on $^4He$; $K^- ~ ^4He \to \Lambda n d$ channel is shown by shadowed area. The figure is taken from \cite{Katz:1970ng}. We added
two arrows indicating the average momentum of the $\Lambda$ for the  two body
absorption, about 550 MeV/c (a), and for the three body absorption, about 650 MeV/c (b).}
\label{new}
\vspace{-0.0cm}
\end{figure}

We start from  the yield of 
3.5 $\pm$ 0.2 \% quoted in \cite{Roosen:1981wu} for the channel $K^- ~ ^4He \to \Lambda n d$, but from only 
two nucleon $K^-$
absorption, since it was guaranteed that the produced deuteron was a 
spectator. 
  The estimated yield does not give yet any information on three body absorption.
Indeed, two situations can be envisaged for the $K^- ~^4He \to \Lambda n d$ reaction: 
two body absorption
$K^-pn \to \Lambda n$, which produces a slow (spectator) deuteron, 
\cite{Roosen:1981wu},
 or three body
absorption  $K^-ppn \to \Lambda d$, which leaves a neutron as a spectator. In this
latter case the deuteron would be produced basically back to back with the
$\Lambda$ and would have a relatively large momentum. In order to get the
strength of the three body absorption process we need extra information, which can be
found by looking 
to the $\Lambda$ momentum spectrum for the 
$K^- ~^4He \to \Lambda d n$ reaction shown in Fig. 2a of
\cite{Katz:1970ng}, and which we reproduce in Fig.   \ref{new}.  In the figure we have
inserted two arrows indicating the average momentum of the $\Lambda$ for the two
situations described before which are about 550 MeV/c for the two body
absorption and 650 MeV/c for the three body absorption. Obviously the 
three body absorption case is penalized dynamically for two reasons: the three body
absorption amplitude should be smaller than the two body one, and forming a $d$
from an excited $\Lambda n p$ system after three body absorption should  be more
difficult than forming a $d$ from a spectator  $n p$ system in the case of
 two body absorption.  The experimental distribution, with
admitted poor statistics, is still significant in as much as it shows strength
below the peak for two body absorption. In order to have such events we must
invoke some extra collision of the $\Lambda$ with the remnant two body spectator
following the dominant two body absorption, which would remove energy from the
$\Lambda$ leading to smaller $\Lambda$ momenta.  The
small experimental bump around 650 MeV/c for the $\Lambda$ momentum should be then 
attributed to the three body absorption process. 

We can make a rough estimate of $8/51$
events  for three body absorption to the total $K^- ~^4He \to \Lambda n d$
yield, or $8/43$ for the  three body to two body absorption ratio. This, together 
with the result from \cite{Roosen:1981wu}, gives us a rate of 
0.65\% for the three body absorption, with large 
uncertainties from the poor statistics of the $\Lambda$ momentum spectrum of 
\cite{Katz:1970ng} (of the order of 40\% from the counts reported). This number is also consistent with the rate of about 1\% 
provided for the
$K^- ~^4He \to \Lambda n d$ reaction with high momentum deuterons in 
 \cite{Suzuki:2007kn}, interpreted there as indicative of three body absorption.
 
 We are aware that $^6Li$ is different from $^4He$ and the
rates could vary from one nucleus to the other. Among other possibilities, 
the deuteron breakup in the final state in $^6Li$ could reduce this rate
somehow.
 Therefore, the qualitative estimate that we have made for the three body
absorption based upon the experimental data of \cite{Roosen:1981wu,Katz:1970ng} in $^4He$
agrees qualitatively with the yield of $0.44\%\ \pm\ 0.14\%$ provided for the peak of the \cite{:2007ph}
experiment, which we have attributed to three body absorption in our kinematical
study of the former sections.  

\section{Conclusions}

We have performed a detailed study of the kinematics following three-nucleon 
$K^-$ absorption in $^6$Li to reanalyze a recent FINUDA experiment, which
claims the existence of a deeply bound kaonic state from the observation of 
a peak in the $\Lambda d$ invariant mass distribution.

Since we are  looking at an inclusive process, where a detailed knowledge on
the nuclear structure is not needed, we have used a local Fermi sea model for
the nucleus, which includes the following basic features: 1) the nucleus is
represented by a local Fermi sea, 2) a kaon from the atomic $2p$ orbit of
$^6$Li is allowed to be absorbed by three nucleons with momenta chosen randomly
within the local Fermi sea, 3) these nucleons are bound by a Thomas Fermi
potential with an additional gap energy chosen to respect the threshold of the
reaction, 4) the total momentum and energy of the initial $K^- NNN$ system is
given by the sum of the individual nucleons and the antikaon, which has zero
three momentum, and 5) this initial $K^- NNN$ system converts into a  $\Lambda
d$ pair and the corresponding value of the $\Lambda d$ invariant mass is
completely determined. The compilation of events provided the distribution
of $\Lambda d$ invariant mass, as well as the momentum
distributions of the individual $\Lambda$ and $d$ in the laboratory frame.

We have been able to reproduce all the basic
features observed in the experiment of Ref.~\cite{:2007ph}, namely the 
invariant mass distribution of $\Lambda d$ pairs, 
the highly correlated back-to-back angular distribution between
the $\Lambda$ and the $d$, the distribution of missing mass with respect to a
final $n d$ system, apart from the measured $\Lambda d$ pair, and the 
momentum distributions of the individual
$\Lambda$ and $d$, as well as that of the combined $\Lambda d$ pair.
In particular, the study served to show that the peak in the  $\Lambda d$ 
invariant mass
distribution observed in the experiment could be naturally reproduced
within a three-nucleon $K^-$ absorption mechanism, 
thus concluding that this observation cannot be used as an
evidence for the existence of a $\bar{K}$ bound on a tribaryon, as was done in 
\cite{:2007ph}. 

    On the positive side, the exercise  served to go one step forward in the
understanding of the process of kaon absorption in nuclei, in this case looking
at the three body mechanism. This interesting phenomenon  deserves a special
attention by itself. Looking at the amount of work that was invested in the
understanding of pion absorption in nuclei, both theoretical and experimental 
\cite{Oset:1986yi,Weyer:1990ye}, 
we can only be satisfied to see that, even if some experiments have been done 
for reasons which could not be supported a posteriori, they
are paving the road for gradually achieving a more complete understanding of the 
phenomenon
of kaon absorption in nuclei, which is necessary for progress in the same search
for possible deeply bound kaon states.

\section{Acknowledgments}
This work is partly supported by contracts FIS2006-03438 and
FIS2005-03142 from MEC (Spain) and FEDER, the Generalitat de
Catalunya contract 2005SGR-00343 and by the Generalitat Valenciana. 
This research is part of the EU
Integrated Infrastructure Initiative Hadron Physics Project under
contract number RII3-CT-2004-506078, and V.K.M.
 wishes to acknowledge direct support from it.


\begin{thebibliography}{0}
\bibitem{weise}
  N.~Kaiser, P.~B.~Siegel and W.~Weise,
  Nucl.\ Phys.\  A {\bf 594} (1995) 325
  [arXiv:nucl-th/9505043].


\bibitem{angels}
  E.~Oset and A.~Ramos,
  Nucl.\ Phys.\ A {\bf 635}, 99 (1998).

\bibitem{Oller:2000fj}
  J.~A.~Oller and U.~G.~Meissner,
  Phys.\ Lett.\ B {\bf 500}, 263 (2001).

\bibitem{Oset:2001cn}
  E.~Oset, A.~Ramos and C.~Bennhold,
  Phys.\ Lett.\ B {\bf 527}, 99 (2002)
  [Erratum-ibid.\ B {\bf 530}, 260 (2002)].




\bibitem{jido}
  D.~Jido, J.~A.~Oller, E.~Oset, A.~Ramos and U.~G.~Meissner,
  Nucl.\ Phys.\ A {\bf 725}, 181 (2003).

\bibitem{Garcia-Recio:2002td}
  C.~Garcia-Recio, J.~Nieves, E.~Ruiz Arriola and M.~J.~Vicente Vacas,
  Phys.\ Rev.\ D {\bf 67}, 076009 (2003).

\bibitem{Garcia-Recio:2003ks}
  C.~Garcia-Recio, M.~F.~M.~Lutz and J.~Nieves,
  Phys.\ Lett.\ B {\bf 582}, 49 (2004).

\bibitem{Hyodo:2002pk}
  T.~Hyodo, S.~I.~Nam, D.~Jido and A.~Hosaka,
  Phys.\ Rev.\ C {\bf 68}, 018201 (2003).

\bibitem{GarciaRecio:2005hy}
  C.~Garcia-Recio, J.~Nieves and L.~L.~Salcedo,
  Phys.\ Rev.\  D {\bf 74} (2006) 034025
  [arXiv:hep-ph/0505233].

\bibitem{Hyodo:2007jq}
  T.~Hyodo and W.~Weise,
  arXiv:0712.1613 [nucl-th].


\bibitem{Beer:2005qi}
  G.~Beer {\it et al.}  [DEAR Collaboration],
  Phys.\ Rev.\ Lett.\  {\bf 94} (2005) 212302.

\bibitem{Borasoy:2005ie}
  B.~Borasoy, R.~Nissler and W.~Weise,
  Eur.\ Phys.\ J.\ A {\bf 25}, 79 (2005).

\bibitem{Oller:2005ig}
  J.~A.~Oller, J.~Prades and M.~Verbeni,
  Phys.\ Rev.\ Lett.\  {\bf 95}, 172502 (2005).

\bibitem{Oller:2006jw}
  J.~A.~Oller,
  Eur.\ Phys.\ J.\ A {\bf 28}, 63 (2006).

\bibitem{borasoy}
  B.~Borasoy, U.~G.~Meissner and R.~Nissler,
  Phys.\ Rev.\  C {\bf 74} (2006) 055201
  [arXiv:hep-ph/0606108].

\bibitem{lutz}
  M.~Lutz,
  Phys.\ Lett.\ B {\bf 426} (1998) 12
  [arXiv:nucl-th/9709073].

\bibitem{angelsself}
  A.~Ramos and E.~Oset,
  Nucl.\ Phys.\ A {\bf 671} (2000) 481
  [arXiv:nucl-th/9906016].

\bibitem{schaffner}
  J.~Schaffner-Bielich, V.~Koch and M.~Effenberger,
  Nucl.\ Phys.\ A {\bf 669} (2000) 153
  [arXiv:nucl-th/9907095].

\bibitem{galself}
  A.~Cieply, E.~Friedman, A.~Gal and J.~Mares,
  Nucl.\ Phys.\ A {\bf 696} (2001) 173
  [arXiv:nucl-th/0104087].

\bibitem{Tolos:2006ny}
  L.~Tolos, A.~Ramos and E.~Oset,
  Phys.\ Rev.\  C {\bf 74} (2006) 015203
  [arXiv:nucl-th/0603033].


\bibitem{okumura}
  S.~Hirenzaki, Y.~Okumura, H.~Toki, E.~Oset and A.~Ramos,
  Phys.\ Rev.\ C {\bf 61} (2000) 055205.

\bibitem{friedman-gal}
E. Friedman, A. Gal, and C. J. Batty, Nucl. Phys. A {\bf 579} (1994) 518 

\bibitem{baca}
A.~Baca, C.~Garcia-Recio and J.~Nieves,
Nucl.\ Phys.\ A {\bf 673} (2000) 335
[arXiv:nucl-th/0001060].


   

\bibitem{gal1}
  E.~Friedman, A.~Gal and J.~Mares,
  Phys.\ Rev.\  C {\bf 60} (1999) 024314
  [arXiv:nucl-th/9804072].

\bibitem{gal2}
  J.~Mares, E.~Friedman and A.~Gal,
  Phys.\ Lett.\  B {\bf 606} (2005) 295
  [arXiv:nucl-th/0407063].

\bibitem{gal3}
  J.~Mares, E.~Friedman and A.~Gal,
  Nucl.\ Phys.\  A {\bf 770} (2006) 84
  [arXiv:nucl-th/0601009].

\bibitem{gal4}
  D.~Gazda, E.~Friedman, A.~Gal and J.~Mares,
  arXiv:0708.2157 [nucl-th].

\bibitem{gal}
  E.~Friedman and A.~Gal,
  arXiv:0705.3965 [nucl-th].
  
\bibitem{muto} T.~Muto, {\it Structure of multi-antikaonic nuclei in the
relativistic mean field model}, talk given at the ``Chiral 07" workshop,
Osaka, Nov. (2007) (http://www.rcnp.osaka-u.ac.jp/~chiral07/).


\bibitem{amigo1}
  T.~Maruyama, T.~Tatsumi, T.~Endo and S.~Chiba,
  Recent Res. Devel. Physics 7 (2006) 1 [arXiv:nucl-th/0605075].

\bibitem{amigo2}
  T.~Maruyama, T.~Tatsumi, D.~N.~Voskresensky, T.~Tanigawa, T.~Endo and S.~Chiba,
  Phys.\ Rev.\  C {\bf 73} (2006) 035802
  [arXiv:nucl-th/0505063].
  
\bibitem{Zhong:2004wa}
  X.~H.~Zhong, L.~Li, C.~H.~Cai and P.~Z.~Ning,
  Commun.\ Theor.\ Phys.\  {\bf 41} (2004) 573.
 
  
\bibitem{Zhong:2006hd}
  X.~H.~Zhong, G.~X.~Peng, L.~Li and P.~Z.~Ning,
  Phys.\ Rev.\  C {\bf 74} (2006) 034321.
  
\bibitem{Dang:2007ai}
  L.~Dang, L.~Li, X.~H.~Zhong and P.~Z.~Ning,
  Phys.\ Rev.\  C {\bf 75} (2007) 068201
  [arXiv:nucl-th/0701060].
    
\bibitem{Torres:2007rz}
  A.~M.~Torres, K.~P.~Khemchandani and E.~Oset,
  arXiv:0712.1938 [nucl-th].

\bibitem{akaishi}
  Y.~Akaishi and T.~Yamazaki,
  Phys.\ Rev.\  C {\bf 65} (2002) 044005.
  
\bibitem{akainew}
  Y.~Akaishi, A.~Dote and T.~Yamazaki,
  Phys.\ Lett.\  B {\bf 613} (2005) 140
  [arXiv:nucl-th/0501040].
  

  
  




\bibitem{toki}
  E.~Oset and H.~Toki,
  Phys.\ Rev.\  C {\bf 74} (2006) 015207
  [arXiv:nucl-th/0509048].
  
\bibitem{akanuc}
  T.~Yamazaki and Y.~Akaishi,
  Nucl.\ Phys.\  A {\bf 792} (2007) 229
  [arXiv:nucl-ex/0609041].
   
\bibitem{hyper}
  E.~Oset, V.~K.~Magas, A.~Ramos and H.~Toki,
proceedings of the IX International Conference on Hypernuclear and
Strange Particle Physics, Mainz (Germany), October 10-14, 2006. Edited by J.
Pochodzalla and Th. Walcher,
(Springer, Germany, 2007), 263
  [arXiv:nucl-th/0701023].


\bibitem{npangels}
A. Ramos, V.K. Magas, E. Oset, H. Toki, Nucl.\ Phys.\ A,  
{\bf doi:10.1016/j.nuclphysa.2008.01.019}.   

  
\bibitem{Shevchenko:2006xy}
  N.~V.~Shevchenko, A.~Gal and J.~Mares,
  Phys.\ Rev.\ Lett.\  {\bf 98} (2007) 082301
  [arXiv:nucl-th/0610022].
  
\bibitem{Shevchenko:2007zz}
  N.~V.~Shevchenko, A.~Gal, J.~Mares and J.~Revai,
  Phys.\ Rev.\  C {\bf 76} (2007) 044004
  [arXiv:0706.4393 [nucl-th]].
  
\bibitem{Ikeda:2007nz}
  Y.~Ikeda and T.~Sato,
  Phys.\ Rev.\  C {\bf 76} (2007) 035203
  [arXiv:0704.1978 [nucl-th]].
  
\bibitem{Yamazaki:2007cs}
  T.~Yamazaki and Y.~Akaishi,
  Phys.\ Rev.\  C {\bf 76} (2007) 045201
  [arXiv:0709.0630 [nucl-th]].
  
\bibitem{Weise:2007rd}
W. Weise, proceedings of the IX International Conference on Hypernuclear and
Strange Particle Physics, Mainz (Germany), October 10-14, 2006. Edited by J.
Pochodzalla and Th. Walcher,
(Springer, Germany, 2007), 243
  [arXiv:nucl-th/0701035].

  
\bibitem{Dote:2007rk}
  A. Dot\'e and W. Weise, proceedings of the IX International Conference on Hypernuclear and
Strange Particle Physics, Mainz (Germany), October 10-14, 2006. Edited by J.
Pochodzalla and Th. Walcher,
(Springer, Germany, 2007), 249
  [arXiv:nucl-th/0701050].

\bibitem{Suzuki:2004ep}
  T.~Suzuki {\it et al.},
  Phys.\ Lett.\  B {\bf 597} (2004) 263.
  
\bibitem{agnellonuc}  
 M.~Agnello {\it et al.}
    [FINUDA Collaboration],
  Nucl.\ Phys.\ A {\bf 775} (2006) 35.

\bibitem{Sato:2007sb}
  M.~Sato {\it et al.},
    Phys.\ Lett.\  B {\bf 659} (2008) 107
  [arXiv:0708.2968 [nucl-ex]].

  
\bibitem{Agnello:2005qj}
  M.~Agnello {\it et al.}  [FINUDA Collaboration],
  Phys.\ Rev.\ Lett.\  {\bf 94} (2005) 212303.
  
\bibitem{Magas:2006fn}
  V.~K.~Magas, E.~Oset, A.~Ramos and H.~Toki,
  Phys.\ Rev.\  C {\bf 74} (2006) 025206
  [arXiv:nucl-th/0601013].
  
 \bibitem{Kpp}
  V.~K.~Magas, E.~Oset, A.~Ramos and H.~Toki,
  [arXiv:nucl-th/0611098];
  A.~Ramos, V.~K.~Magas, E.~Oset and H.~Toki,
  Eur.\ Phys.\ J.\  A {\bf 31} (2007) 684
%
  [arXiv:nucl-th/0702019].
  %
  
\bibitem{:2007ph}
  M.~Agnello {\it et al.}  [FINUDA Collaboration],
  Phys.\ Lett.\  B {\bf 654} (2007) 80
  [arXiv:0708.3614 [nucl-ex]].
  

  
\bibitem{Suzuki:2007kn}
  T.~Suzuki {\it et al.}  [KEK-PS E549 Collaboration],
  Phys.\ Rev.\  C {\bf 76} (2007) 068202
  [arXiv:0709.0996 [nucl-ex]].
  
\bibitem{Katz:1970ng}
  P.~A.~Katz, K.~Bunnell, M.~Derrick, T.~Fields, L.~G.~Hyman and G.~Keyes,
  Phys.\ Rev.\  D {\bf 1} (1970) 1267.
  
\bibitem{Oset:1986yi}
  E.~Oset, Y.~Futami and H.~Toki,
  Nucl.\ Phys.\  A {\bf 448} (1986) 597.
  
\bibitem{harry} T.H.S Lee and K. Ohta, Phys. Rev. Lett. 49 (1982) 1079
  
\bibitem{Weyer:1990ye}
  H.~J.~Weyer,
  Phys.\ Rept.\  {\bf 195} (1990) 295.
  

  
\bibitem{Salcedo:1987md}
  L.~L.~Salcedo, E.~Oset, M.~J.~Vicente-Vacas and C.~Garcia-Recio,
  Nucl.\ Phys.\  A {\bf 484} (1988) 557.
  
\bibitem{Kosmas:1996fh}
  T.~S.~Kosmas and E.~Oset,
  Phys.\ Rev.\  C {\bf 53} (1996) 1409.
  
\bibitem{Albertus:2001pb}
  C.~Albertus, J.~E.~Amaro and J.~Nieves,
  Phys.\ Rev.\ Lett.\  {\bf 89} (2002) 032501
  [arXiv:nucl-th/0110046].
  
\bibitem{Nieves:2004wx}
  J.~Nieves, J.~E.~Amaro and M.~Valverde,
  Phys.\ Rev.\  C {\bf 70} (2004) 055503
  [Erratum-ibid.\  C {\bf 72} (2005) 019902]
  [arXiv:nucl-th/0408005].
  
\bibitem{Roosen:1981wu}
  R.~Roosen and J.~H.~Wickens,
  Nuovo Cim.\  A {\bf 66} (1981) 101.

\end{thebibliography}
\end{document}